\documentclass[prl,twocolumn,showpacs]{revtex4}

\usepackage{graphicx}
\usepackage{amsmath}
\usepackage{amssymb}

\begin{document}
\def\a{\alpha}
\def\b{\beta}
\def\e{\varepsilon}
\def\d{\delta}
\def\l{\lambda}
\def\m{\mu}
\def\t{\tau}
\def\n{\nu}
\def\o{\omega}
\def\r{\rho}
\def\S{\Sigma}
\def\G{\Gamma}
\def\D{\Delta}
\def\O{\Omega}

\def\ra{\rightarrow}
\def\ua{\uparrow}
\def\da{\downarrow}
\def\pd{\partial}
\def\bk{{\bf k}}
\def\bp{{\bf p}}
\def\bn{{\bf n}}

\def\be{\begin{equation}}
\def\ee{\end{equation}}
\def\bea{\begin{eqnarray}}
\def\eea{\end{eqnarray}}
\def\nn{\nonumber}
\def\lb{\label}
\def\pref#1{(\ref{#1})}


\title{Effects of extended impurity perturbation in d-wave 
superconductor}

\author{Yu.G.~Pogorelov and M.C.~Santos}
\affiliation{CFP/Departamento de F\'{i}sica, Universidade do Porto, 
4169-007 Porto, Portugal}

\date{\today }

\begin{abstract}
We describe the effects of electronic perturbation distributed on nearest
neighbor sites to the impurity center in a planar \textit{d}-wave
superconductor, in approximation of circular Fermi surface. Alike the
behavior previously reported for point-like perturbation and square Fermi
surface, the quasiparticle density of states $\rho (\varepsilon )$ can display
a resonance inside the gap (and very weak features from low symmetry 
representations of non-local perturbation) and asymptotically vanishes at 
$\varepsilon \rightarrow 0$ as $\rho\sim\varepsilon/\ln^2\varepsilon$. 
The local suppression of SC order parameter in this model is found to be 
somewhat weaker than for an equivalent point-like (non-magnetic) perturbation 
and much weaker than for a spin-dependent (extended) perturbation.
\end{abstract}

\pacs{74.25.Jb, 74.62.Dh, 74.72.-h}

\maketitle

\section{Introduction}

The study of the density of states (DOS) in high-T$_{c}$ superconducting (SC)
metal oxides has motivated many theorists and experimentalists through the
last years, because it defines such fundamental physical parameters as the
quasiparticle conductivity $\sigma $, the penetration length $\lambda $, the
electronic specific heat, etc. This study is guided by the facts that i) the 
charge carriers are practically confined to the CuO$_{2}$ planes \cite{gin} 
and characterized by 2D wavevectors $\mathbf{k}=(k_{x},k_{y})$  and ii) 
the SC order parameter has \textit{d}-wave symmetry \cite{tsuei} with four 
nodal points $\mathbf{k}_{i}=(\pm k_{\mathrm{F}}/\sqrt{2},\pm k_{\mathrm{F}}/
\sqrt{2})$, where the Fermi surface crosses the nodal directions $k_{x}=
\pm k_{y}$, the SC gap function $\Delta _{\mathbf{k}}$ turns zero, and the 
quasiparticle dispersion law $E_{\mathbf{k}}=\sqrt{\xi _{\mathbf{k}}^{2}+
\Delta _{\mathbf{k}}^{2}}$ coincides with that of normal metal, $\xi _
{\mathbf{k}}$. Another important factor is that the high-T$_{c}$ materials 
are the so-called doped metals, where the Fermi energy $\varepsilon _{\mathrm{F}}$ 
is defined by the density of charge carriers, introduced by the doping process 
\cite{and}. This very process creates the scattering centers for quasiparticles, 
due to random Coulomb fields from ionized dopants. Other scatterers, not related 
to the density of carriers, can be additionally introduced into the system, and 
all them can produce considerable effects on its physical properties \cite{lok01}.

In particular, some resonances can emerge in the quasiparticle spectrum with 
\textit{d}-wave gap symmetry \cite{bal,pog}, even at low concentrations of
impurities. Such resonance manifests itself in a maximum of DOS at a certain
energy $\varepsilon _{res}<\Delta =\max_{\mathbf{k}}\left| \Delta _{\mathbf{k%
}}\right| $, as well as in logarithmic suppression of DOS at $\varepsilon
\ll \varepsilon _{res}$ \cite{lok} (the energies being referred to 
$\varepsilon_{\rm F}$). These conclusions can be directly compared to 
experimental results, as those obtained in the scanning tunnelling microscopy 
experiments \cite{pan}.

The perturbation that impurities introduce into the electronic subsystem of
crystal, depend either on their positions with respect to the lattice and on
the potential they produce on nearest matrix sites. Within the simplest
possible model, where an impurity only disturbs a single site in the
lattice \cite{lee,bal,pog,franz,atk,polk}, the potential is characterized by 
a single perturbation parameter. This point-like perturbation model allows one 
to obtain a simple solution for the quasiparticle DOS in terms of their Green 
functions, leading to the above mentioned possibility of low energy resonances. 
However, in reality, the impurity perturbations in high-T$_{c}$ materials are 
not exactly point-like but rather extended to a finite number of nearest neighbor
lattice sites to the impurity center. This raises an important question on
how robust are the results of point-like approximation to the spatial extent
and geometry of impurity perturbation. The opposite limit to the point-like 
perturbation, when the defect is much bigger of the Fermi wavelength 
and can be treated quasiclassically \cite{ada}, hardly applies to real atomic 
substitutes in high-T$_c$ systems where the perturbation extends to few nearest 
neighbors of the impurity site. Recently, an example of such extended impurity
center was considered for a specific type of spin-dependent perturbation 
\cite{mnm}. Here we use a similar approach to get comparison of the point-like 
and extended perturbations within the same type of perturbation operator, 
impossible for the above mentioned case.

Usually, the treatment of low energy excitations in \textit{d}-wave systems
involves a certain parametrization of the spectrum in the vicinity of nodal
points. Thus, in the popular approach proposed by P. Lee \cite{lee}, one
expands the difference $\mathbf{k}-\mathbf{k}_{i}$ in the local axes, 
$\mathbf{e}_{i,1}=\mathbf{k}_{i}/k_{\mathrm{F}}$ and $\mathbf{e}_{i,2}=
(\mathbf{e}_{i+1,1}-\mathbf{e}_{i-1,1})/2$ (Fig. \ref{fig1}a), and then 
approximates the spectrum components as:
\begin{equation*}
\xi _{\mathbf{k}}=\hbar v_{\mathrm{F}}k_{1},\ \ \ \ \ \Delta _{\mathbf{k}%
}=\hbar v_{2}k_{2},  
\end{equation*}
with two characteristic velocities $v_{2}\ll v_{\mathrm{F}}$.
\begin{figure}
\centering{
\includegraphics[width=9 cm, angle=0]{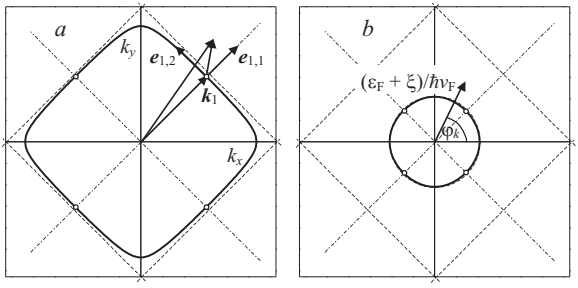}}
\caption{Fermi surfaces for a $d$-wave superconductor. a) ''Square-like'' 
geometry at closeness to half-filling ($\varepsilon_{\rm F}/W = 0.475$). The local 
axes $e_{1,1}$ and $e_{1,2}$ are shown explicitly for the wave vector 
$\mathbf{k}$ near the nodal point $\mathbf{k}_{1}$, in the rest of quadrants 
they are related to the respective nodal points in an analogous way. 
b) ''Circular-like'' geometry at lower doping ($\varepsilon_{\rm F}/W=0.125$), 
almost indistinguishable from exact circle (thin dashed line).}
\lb{fig1}
\end{figure}
However, this implies a ''square'' geometry of Fermi surface and, strictly
speaking, it only holds close enough to the half-filling condition $%
\varepsilon _{\mathrm{F}}/W\approx 1/2$ (where $W$ is the bandwidth). Instead, in
many high-T$_{c}$ compounds we have $\varepsilon _{\mathrm{F}}/W \ll 1$ and the
geometry of Fermi surface is closer to circular (Fig. \ref{fig1}b). In this case, 
a more adequate parametrization of spectrum is obtained with $\xi _{\mathbf{k}%
}=\hbar v_{\mathrm{F}}\left( k-k_{\mathrm{F}}\right) $ and $\Delta _{\mathbf{%
k}}=\Delta \cos 2\varphi _{k}\theta (\varepsilon _{\mathrm{D}}^{2}-\xi _{%
\mathbf{k}}^{2})$ where $\varphi _{k}=\arctan (k_{y}/k_{x})$ and the
theta-function factor restricts SC pairing to the BCS shell of width $%
\varepsilon _{\mathrm{D}}$ (Debye energy) around $\varepsilon _{\mathrm{F}}$.

Most of the known treatments of impurity effects in doped and disordered $d$%
-wave SC systems, including a self-consistent T-matrix analysis of DOS, were
developed within point-like perturbation model and ``square'' geometry of
Fermi surface \cite{lok01,lok}. The present paper is aimed to extend those
studies, in order to include a more realistic features, either of the
impurity perturbation (which can affect several equivalent neighbor sites to
the impurity ion) and of the Fermi surface geometry (in a more adequate
circular approximation).

\section{Physical description of the system}

\begin{figure}
\centering{
\includegraphics[width=6 cm, angle=0]{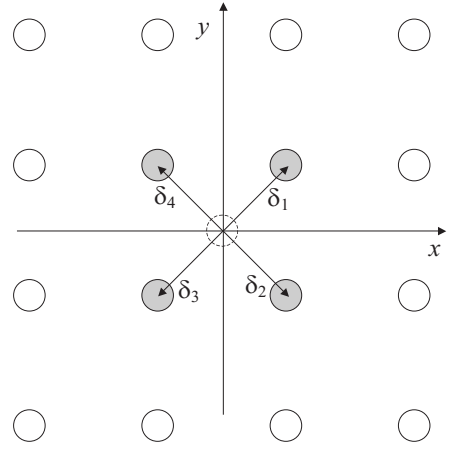}}
\caption{Extended perturbation over four nearest neighbor sites to the impurity 
ion (its projection onto the CuO$_2$ plane is shown by the dashed circle at 
the origin).}
\label{fig2}
\end{figure}

We use the Nambu spinors $\Psi _{\mathbf{k}}^{\dagger}=(a_{\mathbf{k},\uparrow}
^{\dagger},a_{-\mathbf{k},\downarrow })$ where $a_{\mathbf{k},\sigma }^
{\dagger}$ and $a_{\mathbf{k},\sigma }$ are the Fermi operators for 
quasiparticles with wave vector $\mathbf{k}$ and spin $\sigma $ and the 
model Hamiltonian for disordered \textit{d}-wave superconductor: 
\begin{eqnarray}
&&H=H_{0}+H_{imp},  \label{eq2} \\
&&H_{0}=\sum_{\mathbf{k}}\Psi _{\mathbf{k}}^{\dagger }(\xi _{\mathbf{k}}%
\widehat{\tau }_{3}-\Delta _{\mathbf{k}}\widehat{\tau }_{1})\Psi _{\mathbf{k}%
},\quad  \notag \\
H_{imp} &=&-\frac{1}{N}\sum_{\mathbf{k},\mathbf{k}^{\prime },\mathbf{p}}%
\mathrm{e}^{i(\mathbf{k}^{\prime }-\mathbf{k})\mathbf{p}}\sum_{\delta }%
\mathrm{e}^{i(\mathbf{k}^{\prime }-\mathbf{k})\delta }\Psi _{\mathbf{k}%
^{\prime }}^{\dagger }\widehat{V}\Psi _{\mathbf{k}}.  \notag
\end{eqnarray}
In what follows we denote by hats the matrices in Nambu indices, e.g., the
Pauli matrices $\widehat{\tau }_{i}$ and the matrix $\widehat{V}=V_{imp}%
\widehat{\tau }_{3}$ which describes the quasiparticle scattering by
extended (attractive) perturbation $V_{imp}$ around an impurity center 
$\mathbf{p}$, over its near neighbors $\delta $ (Fig. \ref{fig2}). Formally, 
this perturbation only differs by the presence of $\widehat{\tau }_{3}$\ 
factor from that considered in Ref. \cite{mnm}. The concentration of randomly 
distributed centers $c=N^{-1}\sum_{\mathbf{p}}1$ (where $N$ is the number of 
cells) is supposed small, $c\ll 1$.

We define the Green function (GF) matrices as: 
\begin{eqnarray}
\widehat{G}_{\mathbf{k},\mathbf{k}^{\prime }}(\varepsilon )&&= \langle \langle 
\widehat{\Psi}_{\mathbf{k}}|\widehat{\Psi}_{\mathbf{k}^{\prime }}^{\dagger }\rangle
\rangle = \notag \\
&& = i \int_{-\infty }^{0}e^{i(\varepsilon -i0)t}\langle \left\{ {\widehat{\Psi%
}_{\mathbf{k}}(}t{),\widehat{\Psi}_{\mathbf{k}^{\prime }}^{\dagger }(0)}\right\}
\rangle dt,  
\label{eq3}
\end{eqnarray}
where $\langle {\ldots }\rangle $ is the quantum-statistical average with
Hamiltonian, Eq. \ref{eq2} and $\left\{ a{(}t{),}b(0)\right\} $ the
anticommutator of Heisenberg operators. The energy $\varepsilon$ is referred to 
the chemical potenial $\mu$ (identified in what follows with the Fermi energy 
$\varepsilon_{\rm F}$). The observable characteristics follow from the averages 
of products of these operators (at given inverse temperature $\beta $), expressed 
through the corresponding GF's by the spectral theorem: 
\begin{equation}
\langle ab\rangle =\int_{-\infty }^{\infty }\frac {d\varepsilon} 
{\mathrm{e}^{\beta \varepsilon}+1}  {\mathrm{Im}}\langle \langle b|a\rangle 
\rangle _{\varepsilon}.
\label{eq3a}
\end{equation}
For the disordered system with the Hamiltonian, Eq. \ref{eq2}, we calculate 
GF's from the basic equation of motion:
\begin{eqnarray}
\widehat{G}_{\mathbf{k},\mathbf{k}^{\prime }}&=&\widehat{G}_{\mathbf{k}}^{0}
\delta _{\mathbf{k},\mathbf{k}^{\prime }}- \notag \\
&-&\frac{1}{N}\sum_{\mathbf{k}^{\prime \prime },\mathbf{p},j}
\mathrm{e}^{i(\mathbf{k}-\mathbf{k}^{\prime\prime })\mathbf{p}}
\alpha _{j\mathbf{k}}\alpha _{j\mathbf{k}^{\prime \prime}}\widehat{G}_
{\mathbf{k}}^{0}\widehat{V}\widehat{G}_{\mathbf{k}^{\prime \prime},
\mathbf{k}^{\prime }}\,.  
\label{eq4}
\end{eqnarray}
Here $\widehat{G}_{\mathbf{k}}^{0}=(\varepsilon -\xi _{\mathbf{k}}\widehat{\tau}%
_{3}-\Delta _{\mathbf{k}}\widehat{\tau}_{1})^{-1}$, and we expanded the
structural function for impurity scattering in Eq. \ref{eq2} as: 
$\sum_{\delta }e^{i(\mathbf{k}^{\prime }-\mathbf{k})\delta}=\sum_{j=1}^{4}
\alpha _{j\mathbf{k}}\alpha _{j\mathbf{k}^{\prime}}$. The functions 
\begin{eqnarray*}
\alpha _{1,\mathbf{k}} &=&2\cos \frac{ak_{x}}{2}\cos \frac{ak_{y}}{2},\qquad
\alpha _{2,\mathbf{k}}=2\cos \frac{ak_{x}}{2}\sin \frac{ak_{y}}{2}, \notag \\
\alpha _{3,\mathbf{k}} &=&2\sin \frac{ak_{x}}{2}\cos \frac{ak_{y}}{2},\qquad
\alpha _{4,\mathbf{k}}=2\sin \frac{ak_{x}}{2}\sin \frac{ak_{y}}{2},
\end{eqnarray*}
realize irreducible representations of the $C_{4}$ point group ($j=1$ being 
related to \textit{A}-, $j=2,3$ to \textit{E}-, and $j=4$ to 
\textit{B}-representations \cite{cot}) and thus satisfy the orthogonality condition
\begin{equation}
\frac{1}{N}\sum_{\mathbf{k}}\alpha _{j,\mathbf{k}}\alpha _{j^\prime\mathbf{k}}=
\delta _{jj^\prime}.
\label{eq5}
\end{equation}
The impurity effects on quasiparticle spectrum are then naturally classified along 
these representations, alike the known results for magnetic impurities in ferro- 
and antiferromagnetic crystals \cite{iz,ilp}.

\section{T-matrix solutions for the Green functions}

The orthogonality of the $\alpha _{j,\mathbf{k}}$ functions implies that in the 
general solution to Eq. \ref{eq4}:
\begin{equation}
\widehat{G}_{\mathbf{k}}=\left[\left(\widehat{G}_{\mathbf{k}}^0\right)^{-1}
-\widehat{\Sigma }_{\mathbf{k}}\right]^{-1},  \label{eq6}
\end{equation}
the self-energy matrix turns additive in these representations: 
$\widehat{\Sigma }_{\mathbf{k}}=\sum_{j}\widehat{\Sigma }_{j\mathbf{k}}$. Each 
partial term in the latter sum can be given by a specific group expansion (GE), 
like those known for point-like impurity perturbations in normal \cite{iv71} or 
superconducting \cite{lok96,lok01} systems and also for extended perturbations in 
magnetic systems \cite{ilp},
\begin{eqnarray}
\widehat{\Sigma }_{j\mathbf{k}} &=&-c\widehat{T}_{j}\left[ 1-c\widehat{A}%
_{j}-c\widehat{A}_{j}^{2}+\right. \notag \\
&&+ c\sum_{\mathbf{n}\neq 0}\left( \widehat{A}_{j}^{3}\left( n\right) 
\mathrm{e}^{-i\mathbf{kn}}+\widehat{A}_{j}^{4}\left(n\right) \right)\times  \notag \\
&&\left. \times \left( 1-\widehat{A}_{j}^{2}\left( n\right) \right) ^{-1}+
\ldots \right] . 
\label{eq7} 
\end{eqnarray}
Here $\widehat{T}_{j}=\widehat{V}(1+\widehat{V}\widehat{G}_{j})^{-1}$ is the
(renormalized) partial T-matrix, and the matrices $\widehat{A}_{j}\left( n\right)$
represent indirect interactions (in $j$th symmetry channel) between scatterers at 
sites $0$ and $\mathbf{n}$: 
\begin{eqnarray*}
\widehat{A}_{j}\left( n\right)=-\widehat{G}_{j}\left( n\right) 
\widehat{T}_{j}, && \widehat{G}_{j}\left( n\right)=\frac{1}{N}\sum_{\mathbf{k}}
\mathrm{e}^{i\mathbf{kn}}\alpha _{j,\mathbf{k}}^{2}\widehat{G}_{\mathbf{k}}, \\
\widehat{A}_{j} = \widehat{G}_{j}\widehat{T}_{j},\qquad && \qquad \widehat{G}_{j}=
\widehat{G}_{j}\left( 0\right).
\end{eqnarray*}
The sum $\sum_{\mathbf{n}\neq 0}$ in Eq. \ref{eq7} describes all the
processes involving pairs of impurities, it implies averaging in random
impurity configurations and hence runs over all the lattice sites $\mathbf{n}
$, the omitted terms are for triples and more of impurities. This defines a 
generalization of the GE approach for extended impurity centers in superconductors.
If the series in the brackets is restricted to its first term, the self-energy 
matrix $\widehat{\Sigma}_{\mathbf{k}}$ becomes independent of $\mathbf{k}$%
\textsc{:}
\begin{equation}
\widehat{\Sigma}_{\mathbf{k}}\rightarrow \widehat{\Sigma}=-c\sum_{j}\widehat{T}_{j}.
\label{eq9}
\end{equation}
Then, for small enough concentration of impurities, the
renormalization of T-matrices can be neglected and we arrive at
\begin{equation}
\widehat{T}_{j}\rightarrow \widehat{T}_{j}^{0}=\widehat{V}(1+\widehat{G}_{j}^{0}
\widehat{V})^{-1},\quad \widehat{G}_{j}^{0}=\frac{1}{N}\sum_{\mathbf{k}}
\alpha _{j,\mathbf{k}}^{2}\widehat{G}_{\mathbf{k}}^{0}. 
\label{eq10}
\end{equation}
The matrix functions $\widehat{G}_{j}^{0}$ can be expanded in the basis of
Pauli matrices
\begin{equation}
\widehat{G}_{j}^{0}=\rho_0\left(g_{j0}+g_{j1}\widehat{\tau }_{1}-
g_{j3}\widehat{\tau}_{3}\right)
\label{eq11}
\end{equation}
where $\rho _{0}=4/(\pi W)$ is the constant DOS in 2D normal system (the absence 
of $\widehat{\tau }_{2}$ component in Eq. \ref{eq11} is related to the fact that
the gap function $\Delta _{\mathbf{k}}$\ is chosen real). Passing in Eq. \ref{eq10} 
from summation in $\mathbf{k}$ to integration in ''polar'' coordinates 
$\xi _{\mathbf{k}}=\xi$ and $\varphi_{\mathbf{k}}=\varphi$ (Fig. 1b) accordingly 
to the rule:
\begin{equation*}
\frac{1}{N}\sum_{\mathbf{k}}f_{\mathbf{k}}\approx \frac{\rho _{0}}{4\pi }%
\int_{-\mu }^{2/\rho _{0}-\mu }d\xi \int_{0}^{2\pi }d\varphi f\left( \xi
,\varphi \right),
\end{equation*}
we calculate the dimensionless coefficient functions $g_{ji}$. Some of them are 
zero by the symmetry reasons: $g_{11}=g_{41}=0$, the rest can be approximated as:
\begin{eqnarray}
g_{j0}\approx \overline{\alpha_j^2} g_0,\qquad&& g_{j3}\approx 
\overline{\alpha_j^2} g_3,  \notag \\
g_{21}=-g_{31}&\approx&\overline{\alpha_2^2} g_1.
\label{eq11aa}
\end{eqnarray}
Here $\overline{\alpha_j^2}$ are the average values of $\alpha_{j\mathbf{k}}^2$ 
over the Fermi surface: $\overline{\alpha_1^2}\approx 4\left(1 - \omega\right)$, 
$\overline{\alpha_{2,3}^2}\approx 4\omega$, $\overline{\alpha_4^2}\approx 
2\omega^2$, where the band occupation parameter $\omega = \mu/W$ is supposed small, 
in concordance with the chosen circular geometry. The functions $g_0$ and $g_1$ 
are known from the studies of point-like perturbations \cite{per,lp04}:
\begin{eqnarray}
g_0(\varepsilon)&=&\frac{\varepsilon}{4\pi }\int_{-\mu }^{2/\rho _{0}-\mu }
d\xi \int_{0}^{2\pi }\frac {d\varphi}{\varepsilon^2 
-  \xi^2-\Delta^2 \cos^2 2\varphi} \notag \\
&\approx&  \varepsilon \left[1/{\tilde\mu}- \rm{F}_1\left(1-\varepsilon^2/\Delta^2\right)
/\Delta\right], 
\label{11a}
\end{eqnarray}
\begin{eqnarray}
g_1(\varepsilon)&=&\frac{\Delta}{2\pi }\int_{0}^{\varepsilon_{\rm D}}
d\xi \int_{0}^{2\pi }\frac {\cos^2 2\varphi\:  d\varphi }{\varepsilon^2 
- \xi^2-\Delta^2 \cos^2 2\varphi} \notag \\
&\approx& \Delta/\varepsilon_{\rm D} + 2\left[\rm{F}_2\left(1-
\varepsilon^2/\Delta^2\right) \right. \notag \\
&&\qquad\qquad\left.+(\varepsilon^2/\Delta^2)\rm{F}_1\left(1-
\varepsilon^2/\Delta^2\right)\right], 
\label{eq11b}
\end{eqnarray}
they include $\tilde\mu=\mu\left(1-2\omega/\pi\right)\approx\mu$ and the functions 
$\rm{F}_1(z)=\rm{K}(1/z)/\sqrt z$ and $\rm{F}_2(z)=\sqrt z \rm{E}(1/z)$ with full elliptic 
integrals of 1st and 2nd kind $\rm K$ and $\rm E$, having similar behavior with the 
elementary functions obtained within square geometry \cite{lok}. In the same similarity, 
the function $g_3=(4\pi)^{-1}\int_{-\mu}^{2/\rho_0-\mu}\xi d\xi\int_0^{2\pi}d\varphi/
(\varepsilon^2-\xi^2-\Delta^2\cos^2 2\varphi)$ is practically  constant: $g_3 \approx 
\ln \sqrt{\pi/(2\omega) - 1}$, within the relevant energy range $|\varepsilon| \ll W,\mu$. 

Using these results, we readily calculate the partial T-matrices, Eq. \ref{eq10}. 
The most important contribution to $\widehat\Sigma$ comes from the $j=1$ term 
(\textit{A}-representation):
\begin{equation}
\widehat{T}_{1}^0 =\frac{v_A}{\overline{\alpha_1^2}\rho_0} 
\frac{v_A g_0-\widehat{\tau}_3}{D_A}, \label{eq11c}
\end{equation}
where $v_A=\overline{\alpha_1^2} V_{imp}\rho_0 /(1-\overline{\alpha_1^2} 
V_{imp}\rho_0 g_3)$ is the dimensionless perturbation parameter in the 
\textit{A}-channel, and $D_A(\varepsilon) = 1-v_A^2 g_0^2(\varepsilon)$ is the 
energy dependent denominator. In particular, it can produce a low energy resonance 
at $\varepsilon=\varepsilon_{res}$ such that ${\rm Re} D_A(\varepsilon_{res})
=0$, analogous to the above mentioned resonance from point-like impurity center. 
This requires that $v_A$ exceeds some critical value $\approx 2/\pi$. 

The contributions from $j=2,3$ (\textit{E}-representation) are:
\begin{equation}
\widehat{T}_{2,3}^0 =\frac{v_E}{\overline{\alpha_2^2}\rho_0} 
\frac{v_E \left(g_0\mp g_1\widehat{\tau}_1\right)-\widehat{\tau}_3}
{D_E}, \label{eq11d}
\end{equation}
with the respective perturbation parameter $v_E=\overline{\alpha_2^2} V_{imp}\rho_0/
(1-\overline{\alpha_2^2} V_{imp}\rho_0 g_3)$ and denominator $D_E = 1-v_E^2 
\left(g_0^2 - g_1^2\right)$. It is less probable to have a resonance effect in 
this channel at low occupation $\omega \ll 1$, since i) the parameter $v_E$ is 
reduced \textit{vs} the \textit{A}-channel value, and ii) there is a competition 
between ${\rm Re}g_0^2$ and ${\rm Re}g_1^2$ in the denominator $D_E$. 

The \textit{B}-channel contribution ($j=4$) has the same structure as the 
\textit{A}-channel term, Eq. \ref{eq11c}, but with $v_A$ replaced by a strongly 
reduced value $v_B = \overline{\alpha_4^2} V_{imp}\rho_0/(1-\overline{\alpha_4^2}
V_{imp}\rho_0 g_3)$, hence it turns even less important than the \textit{E}-channel 
terms.

\section{Perturbation of observable values}

\begin{figure}
\centering{
\includegraphics[width=8 cm, angle=0]{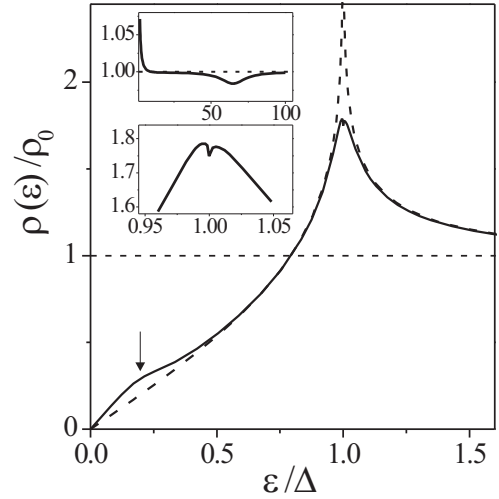}}
\caption{Density of states in the \textit{d}-wave superconductor with extended impurity 
centers (the solid line), for the choice of parameters $W = 2$ eV, $\mu = 0.3$ eV, 
$\varepsilon_{\rm D} = 0.15$ eV, $V_{imp} = 0.2$ eV, $c = 0.15$. The arrow indicates 
the low-energy resonance by the \textit{A}-channel impurity effect and the dashed line 
represents the pure \textit{d}-wave DOS. Insets: weak \textit{E}-channel 
``antiresonances'' at high energies (upper panel) and near the gap edge (lower panel).}
\label{fig3}
\end{figure}

Now we are in a position to describe the perturbation of basic observable 
characteristics of SC system by extended impurity centers. Thus, the 
global DOS is defined by the momentum diagonal GF's, $\rho(\varepsilon)=(\pi N)^{-1} 
\sum_{\mathbf k} {\rm Im\: Tr\:} \widehat{G}_{\mathbf k}$, and, using 
Eqs. \ref{eq9}, \ref{eq11c}, \ref{eq11d} in Eq. \ref{eq6}, it is obtained as
\be
\rho(\varepsilon)=\frac{\rho_0}{\pi}{\rm Im\:} g_0(\varepsilon - \Sigma_0).
\label{eq11e}
\ee
Here the scalar self-energy 
\begin{equation}
\Sigma_0=\frac{c g_0(\varepsilon)}{\rho_0} \left(\frac{v_A^2}
{\overline{\alpha_1^2} D_A}+\frac{2v_E^2}{\overline{\alpha_2^2} D_E} 
+ \frac{v_B^2}{\overline{\alpha_4^2} D_B}\right)
\label{eq11f}
\end{equation}  
includes the effects of extended impurity centers in all three channels. 
The result of direct calculation from Eq. \ref{eq11e} with use of Eq. 
\ref{eq11f} for the characteristic choice of parameters, $W = 2$ eV, 
$\mu = 0.3$ eV, $\varepsilon_{\rm D} = 0.15$ eV, $V_{imp} = 0.2$ 
eV (giving for particular channels: $v_A \approx 0.934$, $v_E 
\approx 0.088$, and $v_B \approx 0.006$), and $c = 0.15$, is shown in 
Fig. \ref{fig3}. It is quite similar to the known results for point-like 
impurities \cite{bal, pog}, showing a reduction of the sharp coherence 
peak at $\varepsilon = \Delta$ and emergence of a relatively broad 
low-energy resonance at $\varepsilon_{res}$ (shown by the arrow), mainly 
due to the \textit{A}-channel effect. But, additionally, there are small 
``antiresonance'' effects from the \textit{E}-channel (insets to Fig. \ref{fig3}), 
at $\varepsilon \approx \Delta$ and at some high enough energy ($\sim 70 \Delta$ 
in this case). Clearly, these \textit{E}-channel features shouldn't have any 
practical effect on the system thermodynamics.  

The local density of states (LDOS) on $\mathbf{n}$th site is expressed in 
terms of GF's as $\rho _{\mathbf{n}}(\varepsilon ) = (\pi N)^{-1}\sum_{\mathbf{k},
\mathbf{k}^{\prime}} \text{Im  Tr  e}^{i (\mathbf{k}-\mathbf{k}^{\prime})\cdot 
\mathbf{n}} \widehat{G}_{\mathbf{k},\mathbf{k}^{\prime}}$ and its variation $\delta 
\rho _{\mathbf{n}}(\varepsilon )=\rho _{\mathbf{n}}(\varepsilon )-
\rho (\varepsilon )$, compared to the mean value $\rho(\varepsilon)=N^{-1}
\sum_{\mathbf{n}}\rho _{\mathbf{n}}(\varepsilon)$ (identical to the global DOS), 
is only given by the momentum-nondiagonal GF's: 
\begin{equation}
\delta \rho _{\mathbf{n}}(\varepsilon )=\frac{1}{\pi N}\sum_{\mathbf{k},%
\mathbf{k}^{\prime }\neq \mathbf{k}}\text{Im Tr e}^{i (\mathbf{k}-\mathbf{k}^{\prime})
\cdot \mathbf{n}}\widehat{G}_{\mathbf{k},\mathbf{k^{\prime }}}.  
\label{eq12}
\end{equation}

\begin{figure}
\centering{
\includegraphics[width=8 cm, angle=0]{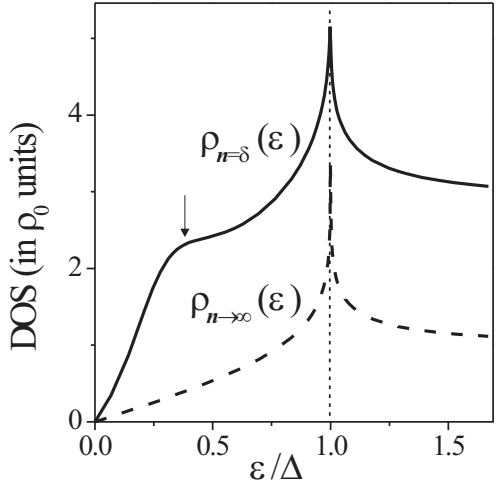}}
\caption{Local density of states on the nearest neighbor site to an 
extended impurity center, for the same choice of parameters as in Fig. 
\ref{fig3} (but supposing $c \to 0$). Note an overall enhancement of 
electronic density compared to remote sites from impurity (dashed line) 
and a much stronger effect of the low-energy resonance (the arrow).}
\label{fig4}
\end{figure}
These functions are easily calculated for the simplest case of a single 
impurity center at ${\bf p}=0$:
\begin{equation}
\widehat{G}_{\mathbf{k},\mathbf{k}^{\prime}}=\frac 1 N \sum _{j}\alpha _{j,\mathbf{k}%
}\widehat{G}_{\mathbf{k}}^0\widehat{T}_{j}^0\widehat{G}_{\mathbf{k}^{\prime}}^0
\alpha _{j,\mathbf{k}^{\prime}},  
\label{eq13}
\end{equation}
describing a finite effect on the local characteristics near the impurity. 
Thus, the quantity $\delta \rho_{\mathbf{n}}$ attains its maximum
value at $\mathbf{n}=\delta$, the nearest neighbor sites to the impurity.
Using Eq. \ref{eq13} and the orthogonality relations, we expand this value in a sum:
\begin{eqnarray*}
\delta \rho_{\mathbf{n} = \mathbf{\delta}}(\varepsilon )=&&\frac{1}{\pi N^{2}}\times \\
 \times \sum_{\mathbf{k},\mathbf{k}^{\prime },j}&&\text{Im  Tr  e}^{i\mathbf{k}
\cdot \delta}\alpha _{j,\mathbf{k}}\widehat{G}_{\mathbf{k}}^0\widehat{T}_{j}^0
\widehat{G}_{\mathbf{k^{\prime }}}^0\alpha _{j,\mathbf{k}^{\prime }}
\text{e}^{-i\mathbf{k^{\prime }}\cdot \delta }= \\
&&=\frac{1}{\pi} \sum_{j}\text{Im Tr} \: \widehat{G}_{j}^0\widehat{T}_{j}^0
\widehat{G}_{j}^0,
\end{eqnarray*}
and present the overall maximum LDOS as:
\begin{eqnarray}
\rho_{\mathbf{n = \delta}}(\varepsilon )&=&\frac{2\rho_0}{\pi}\text{Im}
\left[g_0(\varepsilon) \left(1 + \right. \right. \notag \\
 &+&\left. \left. \frac{v_A N_A}{\overline{\alpha_1^2}D_A} + 2\frac{v_E N_E}
{\overline{\alpha_2^2}D_E} +\frac{v_B N_B}{\overline{\alpha_4^2}D_B}\right)\right].
\label{eq14}
\end{eqnarray}
Alike Eq. \ref{eq11f} for global DOS and the case of Ref. \cite{mnm} for LDOS, 
the resonance contribution to Eq. \ref{eq14} comes from the \textit{A}-channel 
with the numerator $N_A=2 g_3 + v_A (g_0^2 +g_3^2)$, while other channels with 
$N_E=2g_3+v_E(g_0^2-g_1^2-g_3^2)$ and $N_B=2g_3+v_B(g_0^2+g_3^2)$ mainly 
contribute to renormalization of the pure \textit{d}-wave DOS $\rho_d(\varepsilon)=
2/\pi \text{Im} g_0(\varepsilon)$. The calculated from Eq. \ref{eq14} behavior 
of LDOS on nearest neighbor sites to the impurity is shown by solid line in 
Fig. \ref{fig4}. It displays a low energy resonance (the arrow), much 
more pronounced than that in the global DOS, Fig. \ref{fig3}, and an overall 
enhancement compared to the LDOS curve for remote sites from impurity 
$\rho_{\mathbf{n}\to\infty}=\rho_d$ (the dashed line). This picture 
can be compared with the direct experimental measurements of differential 
conductance through the STM tip positioned close to and far from an impurity 
center \cite{pan}.

In a quite similar manner, the local perturbation of SC order parameter can be 
considered. The local \textit{d}-wave SC order in the unit cell containing the 
impurity (Fig. \ref{fig2}) is given by the average $\Delta_{32}=2V\langle 
a_{\delta _{3},\downarrow} a_{\delta _{2},\uparrow }\rangle$ \cite{mnm}, 
where $V$ is the SC coupling constant and site operators $a_{\mathbf{n},\sigma }$ 
are expressed through band operators: $a_{\mathbf{n},\sigma }=N^{-1/2}\sum_
{\mathbf{k}}\text{e}^{i\mathbf{k}\cdot \mathbf{n}}a_{\mathbf{k},\sigma }$. For 
$V_{\text{imp}}=0$, this average coincides with the uniform gap parameter:
\begin{eqnarray} 
&&\quad\Delta=\frac{2V}{N}\sum_{\mathbf{k}}\text{e}^{i\mathbf{k}\cdot(\delta_{2}-
\delta_{3})}\langle a_{-\mathbf{k},\downarrow}a_{\mathbf{k},\uparrow}\rangle\
\notag \\
&=&\frac{4\lambda\Delta}{\pi}\int_0^{\varepsilon_{\rm D}/\Delta}\left[\rm{F}_2(1+x^2)-
\rm{F}_1(1+x^2)\right] d x ,
\label{eq15}
\end{eqnarray}
the latter expression (with the dimensionless \textit{d}-wave coupling constant 
$\lambda = V\rho_0\omega$) being obtained from Eqs. \ref{eq3a} and \ref{eq11b}. 
The integral in Eq. \ref{eq15} behaves as a logarithm: $\int_0^a [{\rm{F}_2}(1+x^2)-
{\rm{F}_1}(1+x^2)] d x \approx(\pi/4)\ln(2.428 a)$ at $a\gg 1$, thus providing 
$\Delta\approx2.428\:\varepsilon_{\rm D}\:\text{exp}(-1/\lambda)$. 

But for $V_{\text{imp}}\neq 0$ this value is locally suppressed. The suppression 
is characterized by the dimensionless parameter $\eta_{sup}=1-\Delta _{32}/\Delta$, 
confined between 0 (pure SC) and 1 (complete local suppression of SC order) \cite{mnm}, 
and it only results from non-diagonal GF's:
\begin{eqnarray}
\eta _{sup}=\frac{2V}{N\Delta }\sum_{\mathbf{k},\mathbf{k}^{\prime }\neq 
\mathbf{k}}\langle a_{-\mathbf{k},\downarrow }a_{\mathbf{k}^{\prime
},\uparrow }\rangle \text{e}^{i(\mathbf{k}\cdot \delta _{2}-\mathbf{k}%
^{\prime }\cdot \delta _{3})}= &&  \notag \\
=\frac{V}{4\pi \Delta }\sum_{j}(-1)^{j}\int_{-\infty }^{0}d\varepsilon {%
\text{Im}\,\text{ Tr}}\widehat{G}_{j}^0\widehat{T}_{j}^0\widehat{G}_{j}^0
\widehat{\tau}_{1}. && 
\label{eq16} 
\end{eqnarray}
Using here Eqs. \ref{eq13} and \ref{eq11} leads to the expression:
\begin{figure}
\centering{
\includegraphics[width=8 cm, angle=0]{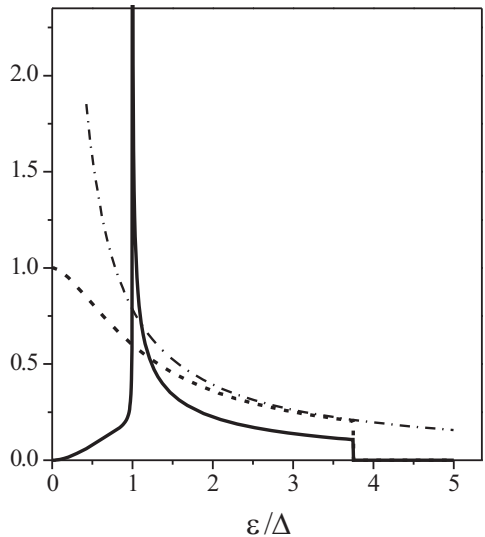}}
\caption{The dimensionless function $F(\varepsilon/\Delta)$ (solid line) used in 
Eq. \ref{eq17} to calculate the suppression parameter $\eta_{sup}$, at the same 
choice of parameters as in Fig. \ref{fig4}, compared to the integrand in the 
uniform gap equation, Eq. \ref{eq15} (dashed line) and its asymptotics 
$\pi\Delta/(4\varepsilon)$ (dash-dotted line).}
\label{fig5}
\end{figure}
\begin{equation}
\eta _{sup}=\frac{4\lambda}{\pi} \int_{0}^{\varepsilon_{\rm D}/\Delta}
F(x)dx,  \label{eq17}
\end{equation}
where only the \textit{E}-channel terms contribute to the dimensionless function 
\begin{equation*}
F(x)= -\frac{v_E}{2} \text{Im}\frac{g_1(x\Delta)N_E}{D_E}
\end{equation*}
(compared in Fig. \ref{fig5} with the integrand for the uniform gap equation, Eq. \ref{eq15}). 
Numeric analysis of this expression for the above chosen perturbation parameters results 
in $\eta _{sup}\approx 0.47$. This is somewhat smaller than the respective value for 
point-like impurity in \textit{d}-wave system \cite{pog}: $\eta_{sup}=1/(1+v^2)$ (assuming
$v$ equal to $v_A$), and more than twice weaker than almost complete suppression in
the case of \textit{spin-dependent} perturbation of the same dimensionless magnitude 
\cite{mnm}. Those relations confirm the general Abrikosov-Gor'kov conclusion \cite{ag} 
on the pair-breaking effects by impurities in superconductors, irrespectively of their 
spatial extension.

\section{Self-consistent generalization}

The above given analysis corresponds to the simplest restriction of the 
group series, Eq. \ref{eq7}, for self-energy to its first, single-impurity 
term with use of unperturbed GF's. The resulting linear approximation in 
impurity concentration: $\widehat\Sigma = -c\sum_j \widehat T_j^{0}$, is only 
justified when this concentration is low enough $c\ll c_{0}$, where $c_{0}\sim 
\rho_0\varepsilon_0$ is related to the characteristic energy scale $\varepsilon_0$ 
for impurity perturbation (in this case $\varepsilon_0\sim\varepsilon_{res}$). 
At higher concentrations, $c > c_0$, when perturbations from different impurity 
centers effectively overlap, a simplest way to take account of these collective 
impurity effects is provided by replacement of $\widehat\Sigma$ by its self-consistent 
analogue $\widehat\Sigma^{(sc)} = -c\sum_j T_j^{(sc)}$, in the spirit of well-known 
self-consistent T-matrix approximation \cite{baym,lee,lok}. It was shown for the 
case of point-like impurity perturbation \cite{lok}, that effects of such 
self-consistency are most essential at the lowest excitation energies, 
$\varepsilon\ll\Delta$. In view of the similarity in the system response to 
point-like and extended perturbations and of the predominant role of the 
\textit{A}-channel at low energies, we can restrict the self-consistency of 
T-matrix only to its \textit{A}-channel term. Then Eqs. (\ref{eq6}) and (\ref{eq9}) 
are modified to:
\begin{equation}
\widehat{G}_{\mathbf{k}}^{(sc)} =\left[\left(\widehat{G}_{\mathbf{k}}^{0}\right)^{-1}
+c\sum_{j=2}^{4}\widehat{T}_j+c\widehat{T}_1^{(sc)}\right] ^{-1},
\label{eq17}
\end{equation}
including the self-consistent \textit{A}-channel T-matrix:
\begin{equation*}
\widehat{T}_{1}^{(sc)}=\widehat{V}[1+\widehat{G}_{1}^{(sc)}\widehat{V}]^{-1}.
\end{equation*}
Alike Eq. \ref{eq11}, the self-consistent GF matrix $\widehat{G}_{1}^{(sc)} = 
N^{-1}\sum_{\mathbf{k}}\alpha _{1\mathbf{k}}^{2}\widehat{G}_{\mathbf{k}}^{(sc)}$ 
can be parametrized in Pauli matrices:
\begin{equation*}
\widehat{G}_{1}^{0}=\rho_0\overline{\alpha_1^2}\left(g-g_{3}
\widehat{\tau}_{3}\right).
\end{equation*}
Then, using Eq. \ref{eq11d} in Eq. \ref{eq17}, we readily conclude that 
$\widehat{\Sigma}^{(sc)}$ is diagonal in Nambu indices that is, within the 
considered self-consistent $T$-matrix approximation for extended impurity 
centers, the scattering by dopants does not influence the $d$-wave order 
parameter, the same as for point-like centers \cite{lok01}. Hence, the 
self-consistency should be achieved only for the scalar function 
$g=(2 N)^{-1}\sum_{\mathbf{k}}\text{Tr}\widehat{G}_{\mathbf{k}}^{(sc)}$ 
through the equation:
\begin{equation}
g(\varepsilon)=g_0(\varepsilon - \Sigma_0^{(sc)}(\varepsilon)),
\label{eq20}
\end{equation}
where the scalar self-consistent self-energy:
\begin{equation}
\Sigma_0^{(sc)}(\varepsilon)=\frac{c g(\varepsilon)}{\rho_0}\left(
\frac{v_A^2}{\overline{\alpha_1^2}D_A^{(sc)}}+\frac{2 v_E^2}
{\overline{\alpha_2^2}D_E}+\frac{v_B^2}{\overline{\alpha_4^2}D_B}\right),
\label{eq21}
\end{equation}
includes the self-consistent denominator: $D_A^{(sc)}(\varepsilon)=
1-v_A^2g^2(\varepsilon)$. Then, passing to dimensionless energy 
$x=\varepsilon/\Delta$ and denoting
\begin{equation}
\frac{\Sigma_0^{(sc]}}{\Delta}=\sigma(g)=\alpha g\left(\frac{\beta}
{1-\alpha^2 g^2}+\beta^{\prime}\right)
\label{eq21a}
\end{equation}
with $\alpha = v_A$, $\beta = c v_A/(\overline{\alpha_1^2}\rho_0\Delta)$, 
$\beta^{\prime}=c[2v_E^2/(\overline{\alpha_2^2}D_E)+v_B^2/(\overline{\alpha_4^2}
D_B)]/(v_A \rho_0\Delta)$ and $\gamma=\Delta/[\mu(1-2\omega/\pi)]$, we arrive at 
the self-consistency equation for $g = g(x)$ as:
\begin{eqnarray}
g&=&\left(x-\sigma\right)\left\{\gamma-\right. \notag \\
&&\left. -\frac{1}{\sqrt{1-\left(x-\sigma\right)^2}}\text{K}\left[\frac{1}{1-
\left(x-\sigma\right)^2}\right]\right\},
\label{eq22}
\end{eqnarray}
with $\sigma = \sigma(g)$ defined by Eq. \ref{eq21a}. This equation is quite 
similar to that reported for point-like impurity centers and square-like 
geometry \cite{lok01, lok}, differing only by the appearance of additional term 
$\beta^\prime$ in Eq. \ref{eq21a}. In the same way, Eq. \ref{eq22} admits 
two types of solutions in the energy range of principal interest $x\to 0$. 
One of them, $g=g^{(1)}(x)$, tends in this limit to a finite imaginary value: 
$g^{(1)}(x\to 0)\to i\gamma_0$, defined by the equation
\begin{eqnarray}
1&=&-\alpha f\left(\gamma_0\right)\left\{\gamma-\frac{1}{\sqrt{1+\alpha^2\gamma_0^2
f^2\left(\gamma_0\right)}}\times \right. \notag\\
&&\left.\times\text{K}\left[\frac{1}{1+\alpha^2\gamma_0^2 f^2\left(\gamma_0\right)}
\right]\right\},
\end{eqnarray}
with $f\left(\gamma_0\right)=\beta^\prime+\beta/(1+\alpha^2 \gamma_0^2)$. 
Another solution, $g=g^{(2)}(x)$, is vanishing in this limit: $g^{(2)}(x\to 0)\to 
0$, so that all the denominators $D$ in Eq. \ref{eq21} can be safely put equal 
unity, simplifying Eq. \ref{eq22} to:
\begin{equation}
g=(x-\alpha^\prime g)\left\{\gamma -\text{K}\left[\frac{1}{1-(x-\alpha^\prime g)^2}
\right]\right\},
\label{eq23}
\end{equation}
with $\alpha^\prime=\alpha(\beta + \beta^\prime)$. Its solution has the same 
logarithmic asymptotics at $x\to 0$:
\begin{equation}
g(x)\approx \frac{x}{\alpha^\prime}\left[1-\frac{1}
{\pi\alpha^\prime\ln(2i\pi\alpha^\prime/x)}\right],
\end{equation}
as found in Refs. \cite{lok01,lok}. Also the conclusion on the validity 
range for each of the two solutions: $g^{(1)}$ far enough from and $g^{(2)}$ 
close to the Fermi level \cite{lp04}, remains true in the present situation. 
Thus the DOS $\rho(\varepsilon)$ at $\varepsilon\ll\varepsilon_{res}$ should 
be suppressed as:
\begin{equation}
\rho(\varepsilon)\approx\rho_0 \frac{\varepsilon}{2\pi\Delta\left[\alpha^\prime
\ln(2\pi\alpha^\prime\Delta/\varepsilon)\right]^2}
\end{equation}
compared to its linear asymptotics $\rho\sim\rho_0\varepsilon/\Delta$ resulting 
from the linear in $c$ approximation, seen in Fig. \ref{fig3}.

\section{Conclusion}
 
The Green function analysis is developed for the quasiparticle spectrum in 
a planar \textit{d}-wave superconductor with finite concentration of impurity centers 
which perturb atomic energy levels on nearest neighbor lattice sites. A 
generalization of the method of group expansions for quasiparticle self-energy 
is obtained for such extended impurity centers. The general picture of spectrum 
restructuring is found quite similar to that previously established for 
point-like impurity perturbation, though some new specific features due to 
extended nature of the perturbation are also indicated. In particular, it is 
found that the effects on the quasiparticle DOS and on the SC order parameter 
result from different irreducible representations of the point symmetry group 
of the impurity center. The self-consistent procedure is proposed for higher 
concentration of extended impurities, generalizing the known formulations for 
point-like centers, and a qualitative similarity with that case is demonstrated.

\end{document}